\documentclass[10pt, preprint2]{aastex}

\begin{document}

\title{Doppler Boosting May Have Played No Significant Role in the Finding Surveys of Radio-Loud Quasars}



\author{M.B. Bell\altaffilmark{1}}

\altaffiltext{1}{Herzberg Institute of Astrophysics,
National Research Council of Canada, 100 Sussex Drive, Ottawa,
ON, Canada K1A 0R6;
morley.bell@nrc-cnrc.gc.ca}

\begin{abstract}

There appears to be a fundamental problem facing Active Galactic Nuclei jet models that require highly relativistic ejection speeds and small jet viewing angles to explain the large apparent superluminal motions seen in so many of the radio-loud quasars with high redshift. When the data are looked at closely it is found that only a small percentage of the observed radio frequency flux density from these sources can be Doppler boosted. Without a highly directed, Doppler boosted component that \em dominates \em the observed flux, radio sources found in low-frequency finding surveys cannot be preferentially selected with small jet viewing angles. The distribution of jet orientations will then follow the sin$i$ curve associated with a random distribution, where only a very few sources ($\sim1\%$) will have the small viewing angles ($<8\arcdeg$) required to explain apparent superluminal motions v$_{app} > 10c$, and this makes it difficult to explain how around $33\%$ of the radio-loud AGNs with high redshift can exhibit such highly superluminal motions. When the boosted component is the dominant one it can be argued that in a flux limited sample only those members with small viewing angles would be picked up while those with larger viewing angles (the un-boosted ones) would be missed. However, this is not the case when the boosted component is small and a new model to explain the high apparent superluminal motions may be needed if the redshifts of high-redshift quasars are to remain entirely cosmological.

\end{abstract}

\keywords{galaxies: active - galaxies: distances and redshifts - galaxies: quasars: general}

\section{Introduction}

Recently \citet{lop11} has reminded us that there are still many quasar/QSO observations that remain difficult to explain. Here we discuss what appears to be another of these. The large apparent superluminal motions observed in the jets of many radio-loud quasars can be explained by assuming either, (a) that the objects are at their cosmological redshift (CR) distance and almost all of their radio flux density comes from ejected material that is relativistically beamed towards us in a highly collimated jet at near light speed and with a small inclination angle, $i$, close to the line-of-sight \citep{ree66,zen87}, or (b) that the objects are much closer than their redshifts imply so the observed angular motions in their jets lead to only subluminal linear speeds \citep{nar89,bel07a}. It has been claimed that the former model not only explains the apparent superluminal motions, but that it can also, through Doppler boosting, explain why most of the detected sources would naturally have very small inclination angles. However, for this model to work, one of its main requirements is that, in the finding survey, the Doppler boosted component of the source flux density must be the dominant one. Whether or not this requirement is met therefore needs to be examined closely. To do this we first examine what source material can be moving towards us at relativistic speeds in a tightly confined beam. We then consider what percentage of the total source flux density the radiation from this material contributes. It will be demonstrated below that with the existing observational evidence it may no longer be possible to use the relativistic beaming model to explain the high percentage of radio loud quasars exhibiting superluminal motion.

\section{The role of Doppler boosting}

The problem of explaining apparent superluminal motion in quasar jets was looked at closely over twenty years ago \citep[see for example]{lin85}. Since that time much new information has been obtained on the jets of many more radio loud quasars. Much of it \citep{kel98,kel04} is of excellent quality, and some of it has resulted in movies being made that depict reasonably clearly what is taking place near the central engines of these objects when ejections occur. Unfortunately the lack of adequate resolution near the central compact object still prevents us from obtaining a clear picture of the jet launching process. If material is ejected from a source at relativistic speeds, because of Doppler boosting its radiation in the direction of motion will be enhanced, and radio finding surveys will preferentially pick up those sources that are ejecting material towards us \citep{kel04}. This is only true, however, as long as the boosted component is the dominant one. The largest boosts in intensity occur for sources with jet viewing angles $i < 8\arcdeg$ \citep[see their Fig 20 and eqn B5]{urr95}. Thus, as pointed out above, if a high percentage of the sources show apparent superluminal motion in their jets it can be explained if most of the radiation has been Doppler boosted and comes from material whose ejection speed is relativistic and whose direction of motion is towards us (close to the line-of-sight). If none, or only a small percentage, of the radiation is Doppler boosted, and there are no other selection effects active, most of the sources would have been detected without the boosted component. The sources will then have a close to random distribution of orientations (sin$i$) in which $50\%$ will have jet viewing angles that are greater than $60\arcdeg$, and only $\sim1\%$ will have $i < 8\arcdeg$. In this case, if a large percentage of the sources show large apparent superluminal motions, another way of explaining these motions must be found. Although they can be explained by bringing the sources closer to us until the linear speeds calculated from their angular motions are no longer superluminal, this argument has been found unacceptable because it requires that the redshifts of quasars contain an additional intrinsic component unrelated to the normal cosmological, or distance-related, one.

\subsection{Jet/Counterjet Asymmetry}

It has been argued that the jet asymmetry, or one-sidedness, seen in many of these objects at 15 GHz, is a strong indication for Doppler boosting in the approaching jet. But recently, an attempt to show that the asymmetry in the jets of M87 at 15 GHz could be explained by relativistic motion gave negative results \citep{kov07}. No evidence was found for relativistic ejections at 15 GHz in spite of the obvious jet/counterjet asymmetry. These authors were forced to conclude that the large jet/counterjet asymmetry in the inner jets of M87 may be intrinsic and not due to Doppler boosting. This was a significant result that appears to negate one of the main claims of the proponents of Doppler boosting; namely that a jet/counterjet asymmetry is evidence for Doppler boosting. Although relativistic motions have been claimed in the M87 inner jet at X-ray wavelengths \citep{bir99}, there is no way to be certain that it is the same material that is being observed at 15 GHz. In fact, it has been suggested that the X-ray event may represent an entirely different phenomenon (see Fig 6 of \citet{bel07b}). The important point here is that when the same observing frequency (15 GHz) was used to observe both the jet motion and the asymmetry, only \em non-relativistic \em ejection speeds were seen in the material that showed asymmetry.

It would seem then that the jet one-sidedness seen in so many of these sources may originate simply because the strength of the jet is associated with the amount of material that is moving from the accretion disc to the jet at any given time, and that this material, associated with the flaring type of jet ejections discussed here, is normally only accreted from one side of the disc at a time. Unfortunately, the exact process by which material is ingested into the central object and regurgitated in the jet is still not well understood. However, the evidence clearly indicates that intrinsic asymmetry is common and its presence then cannot automatically be assumed to imply relativistic beaming in many sources.

\section{What Percentage of the Radiation Comes From Material Moving Outward at Relativistic Speeds?}

We now need to determine what percentage of the total radio radiation from these sources is likely to originate in material that is moving away from the central object at relativistic speeds, and in a tightly confined beam. To help in this examination the different emission regions found in jetted sources are shown in Fig 1. There are three main regions: (1) an extended, kpc-scale jet that is resolved in the VLA observations but lies outside the field of view of the VLBA. Although not shown here, the extended jet may also include at its end a giant radio lobe with hot spots. (2) an inner, parsec-scale jet that is well resolved in VLBA observations but unresolved by the VLA and, (3) a compact core component that is unresolved in all cases. In the original finding surveys that were carried out at long radio wavelengths it is entirely possible that much of the radiation from the inner, pc-scale jet may have originated from a region that was too deep inside the long-wavelength radio photosphere to have been detectable.

\begin{figure}[t]
\hspace{-0.5cm}
\vspace{-2.0cm}
\epsscale{1.0}
\includegraphics[width=9cm]{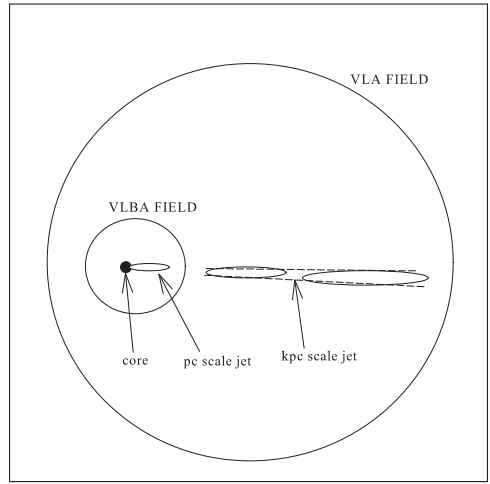}
\caption{{Figure showing relative jet areas covered by VLBA and VLA (not to scale). Because the core is unresolved a small portion of the inner, pc-scale jet flux will be included in the peak flux of the core component. However, because in most cases the inner jet covers several beam areas, this is expected to be small.
\label{fig1}}}
\end{figure}

\subsection{Radiation from the Outer Jet and Giant Radio Lobes}


As noted earlier, when the Doppler boosted component is small compared to the total flux density it cannot introduce a strong selection effect that will preferentially pick up sources with small inclination angles in the finding surveys in which most of these radio-loud AGN galaxies were discovered. Almost all radio-loud AGN galaxies (quasars, BLLacs) were found in the early radio surveys (Parkes, Cambridge 3C and 4C), that were carried out at low frequencies (178 or 408 MHz) with large antenna beamwidths. The beamwidth of the Parkes telescope at 408 MHz, for example, is $48\arcmin$ \citep{bol64}. Consequently, the finding surveys would have detected the total radio radiation coming from these sources. This is especially true for high redshift sources where the largest angular size is less than $\sim3\arcmin$ for sources with $z>0.1$ \citep{mil71}. Even the 4C survey, which had a $1.3\arcmin$ beam, would have detected the total radiation from sources with $z > 0.2$, which includes most of the radio-loud quasars. The 4C detection limit was 2 Jy.

When a jetted source contains giant lobes with internal hotspots these features will almost certainly contain most of the source flux. Since outward motion in the lobes has been shown to be close to 0.02$\pm0.01$c \citep{ale87,cle06,ode09}, the radiation from the lobes will be unboosted.

Deceleration of the flow in the kpc-scale jet has also been examined by several previous investigators \citep{ode09,lai99,lai02a,lai02b,lai06,lai08} and the flow is found to slow down quickly to near 0.1c beyond a few kpc from the core. As a result radiation from most of the kpc-scale jet must then also be un-boosted. Thus none of the radiation from the lobes and almost none of that from the kpc-scale jet can be included in the relativistic flow radiation component. This means that the boosted radiation component must come mainly from the core or inner pc-scale jet.

\subsection{Radiation from the Core and Inner Jet}

The question of whether or not the radiation from the compact central core in core-dominant sources is boosted is obviously an important one. Because of this a lot of effort has been devoted to trying to prove that the core radiation, which is unresolved even with the mas resolution of the VLBA, is boosted in the jet direction. Some investigators have argued that the core component, although stationary, is actually part of the jet base \citep{mar77,mar08,mar09}, and that the lack of core motion in this case is because the material in this region is still being accelerated and has not yet reached relativistic speeds. But if this were the case it would not matter whether the core is associated with the accretion disc or the jet base, its radiation cannot be boosted if the radiating material is not moving relativistically.

Recently it was demonstrated \citep{bel10} that most of the radio frequency radiation from the strong, unresolved cores of these objects could not originate in the jet, and must be coming from a separate region centered on the central compact object and accretion disc. It is apparent \citep{bel10} from the investigation of 3C279 by \citet{cha08} that there are three separate radiation components involved in producing the total radio radiation from the compact core and inner jet. 

These are as follows:

1) The first is a flaring component that only becomes visible when a new ejection event commences, and then only after the radiating material being ejected passes beyond its relevant $photosphere$, which is the point beyond which the external medium is transparent to the wavelength being observed. This component is jet related and at radio frequencies represents only a small percentage of the total flux density observed (comparable to that from an individual blob seen after the ejected material begins to be resolved in the inner jet). For outward motion in the jet, since we see deeper at short wavelengths, the shorter wavelength flares ($\gamma$-ray, X-ray, optical) will appear before the radio flare. It is thus possible to estimate a lower limit to the separation between any two $photospheres$ from the time delay in light days between the appearance of their respective flares. Recently, \citet{jor10} have carried out an analysis of the flaring behavior of 3C454.3 using short wavelength data (optical, X-Ray, $\gamma$-ray), as well as mm-wave. They find a time delay of 30$\pm$15 light days between the short wavelengths and the mm-wave flares. This corresponds to a distance of $\sim$0.025 pc if the jet material is moving out relativistcally. It is generally accepted that the high energy flaring   radiation comes from the unresolved region very close to the accretion disc, and from the 3C454.3 results this appears to be confirmed, with the radius of the mm-wave photosphere likely to be less than 0.1 pc. From these results it then seems likely that the radius of the \em radio photosphere \em lies well inside the half-power beam of the VLBA, even at 43 GHz. Note that in this model there is no \em Blazar zone \em in the jet of the type postulated by \citet{sik08}. For a particular wavelength this zone is replaced in our model by the point in the envelope surrounding the accretion disc at which the jet first becomes visible (its relevant $photosphere$). This point is, unfortunately, still unresolved by present-day instruments. At radio frequencies this flaring component is much weaker than the total core component, is jet related but still unresolved, and is expected to be boosted if the jet is pointing towards us. The visibility of the flaring components increases towards shorter wavelengths and this can be explained by the decrease in radius of the respective photospheres which translates into an increase in the magnetic field strength closer to the central compact object.

2) The second component observed in the flux monitoring of 3C279 is one that can be referred to as the slowly varying component. It is easily shown that this component comes mainly from the inner jet, increasing and decreasing directly with the number of blobs present in the jet. This component is entirely related to the jet material, varies continuously, and if moving in our direction is expected to be Doppler boosted. In the most active sources this component can be as least as strong as the core component. Although the highly variable sources (like 3C279 and 3C454.3) are the most highly studied, the majority of core-dominant sources do not fall into this highly variable category and in most cases the core component is the dominant one.

3) The third component is a non-varying one that contains most of the flux from the unresolved radio core. Since this radiation is detectable, at radio frequencies it must come from a radius larger than that of the radio photosphere. Even with the best resolution available this component is centered on the accretion disc, and shows no sign of motion. There is no evidence that any of the non-varying core component is associated with the jet flow. However, because both it and the flaring component are unresolved, the two will be superimposed, even with the resolution provided by the VLBA.

It was demonstrated \citep{bel10} that the non-flaring core component cannot be explained by a continuous jet flow component, which, if it were part of the jet, would be needed to explain its steady nature.

\citet{jor10} argue that the radio core lies at the end of the acceleration zone at the base of the jet. This can easily be ruled out when there is no continuous flow, because this dominant, non-varying core component is still visible even when there is no ejection event taking place to be accelerated. Furthermore, if it were part of the jet, it is not clear how this strong radio core can show no sign of motion, while motion is readily seen as soon as the material moves outside the photosphere. If the radiation is coming from a region in the jet that is not moving, and produced by particles that are passing through it at relativistic speeds, the radiation from this stationary core material still cannot be boosted. Is no motion seen in the acceleration region because the viewing angle is close to 0$\arcdeg$, while motion is readily detected beyond the core because there is a change in the jet direction at the core? Although it is suggested in their core-in-jet model that there may be a sudden change in the jet direction at the core, the likelihood that every source would have this same bend seems small. It is also interesting that, while no proper motion is seen in the core, relativistic motion in the inner jet is readily detected even though the viewing angle of these components cannot differ by more than a few degrees $(<5\arcdeg)$ if both are to be highly boosted. In the CR model this effectively rules out the possibility of a significant change in direction between the motion in the acceleration zone and the motion further out. It also needs to be kept in mind in this model that the superluminal motion is seen in the portion of the jet that would have the largest viewing angle. It seems very unlikely that this core-in-jet model can be a viable one and our previous conclusions \citep{bel10} that the radio core is un-boosted and centered on the accretion disc remains much more likely.

In fact \citet{hom02} have difficulty explaining the brightness they see in some of the features in the mas jet of PKS 1510-089 by Doppler boosting, arguing that the brightness must be dominated by shocked emission. This is very damning for the relativistic beaming model. Also, as they too admit, the high levels of fractional polarization they detect in the outer edge of the mas jet suggests that the bow shock is seen from the side, which would be the case if the viewing angle of the jet was large as is being suggested here, instead of coming towards us as would be the case in their model.

From the above examination it is concluded here that most of the flux from the core component is un-boosted, with almost all of the boosted radiation in these sources originating then in the inner jet. This conclusion is also consistent with the fact that it is only in the inner jet that apparent superluminal motions have been conclusively detected. We are now interested in determining what percentage of the total flux would have come from the inner jet in the finding survey.

\section{Relative Strengths of the Boosted and Un-boosted Radiation}

In the VLBA contour plots of the core and inner mas jets of 132 radio-loud AGN galaxies (radio galaxies, BLLacs and quasars) obtained at 15 GHz \citep{kel98}, the flux density from the unresolved, compact core component dominates that from the pc-scale, inner jet in most cases. Twenty of these sources, chosen mainly because they have very high $\beta_{app}$ values, are included in Table 1. Here $\beta_{app}$ = v$_{app}$/c, where v$_{app}$ is the apparent linear speed in the jet obtained from the observed angular motion, assuming that the source is located at a distance determined from its redshift. Because of their high apparent superluminal motions the jets of these sources must have very small viewing angles if these motions are to be explained in the relativistically beaming model. From their contour plots it can be seen that the inner, pc-scale jets of these sources almost always cover several beam areas. Because the core component is unresolved it is assumed that a small part of the inner jet component lying at the base of the pc-scale jet would have been included in the peak flux of the core. This can represent only a very small portion of the core flux, however, when the inner jet covers several beam areas, and the entire inner jet radiation component is itself, in most cases, much smaller than the peak core component. 

Thus, although the core and inner end of the jet cannot be resolved, the component of the flux coming from the inner jet but included in the core peak flux will be negligible. From this we have estimated the approximate 2-cm flux from the pc-scale jet, S$_{in}$, using the relation S$_{in}$ = S$_{total}$ - S$_{peak}$, where S$_{total}$ and S$_{peak}$ are flux values obtained with the VLBA and have been taken from Table 3 of \citet{kel98}. S$_{in}$ has been included in column 7 of Table 1. Columns 5 and 6 give the source flux densities measured in the finding surveys at 178 MHz or 400 Mhz, taken from the Dixon catalog \citep{dix70}, the Parkes catalog \citep{eke69}, the 4C(+20$\arcdeg to +40\arcdeg$) catalog \citep{pil64}, and the 4C(-7$\arcdeg to +20\arcdeg$ and +40$\arcdeg to +80\arcdeg$) catalog \citep{gow67}. S$_{ext}$, included in column 8 of Table 1, represents the flux from the external, (kpc), jet component taken from \citet{mur93} and \citet{kha10}.

From the examination carried out in the previous section, for the purposes of this investigation we shall assume, (a) that the radiation from the core, S$_{peak}$, is almost entirely unboosted, (b) that the material in the inner jet is almost certainly to be moving relativistically in the CR model and will be boosted if its direction is towards us, and (c) that most of the material in the external, kpc-scale jet in these core-dominant sources is not moving relativistically and will therefore not be boosted.


In Table 1, column 9 lists F$_{IJ}$, the ratio of the inner jet flux found at 2 cm, where the resolution is adequate to resolve it, to the total flux found at the low frequencies of the finding surveys, expressed as a percentage. We assume here that the spectral index of the jet is flat even though there is a good chance that all, or at least part, of the inner jet may be located inside the low-frequency photosphere, which would prevent its detection at the low radio frequencies used in the finding surveys. If this is the case, the value of F$_{IJ}$ would be even smaller than the value listed.

To be considered dominant F$_{IJ}$ needs to make up more than 90 percent of the total flux. As can be seen in Table 1, no sources come close to this. Even when it is assumed in column 9 that the external kpc jet flux is also boosted, the entire jet component, F$_{EJ}$, is also far from dominating the total source flux. It is therefore not possible for the inner jet, or even the entire jet, to have introduced into the finding survey a strong selection effect that would have preferentially chosen sources with small inclination angles. As noted above, this is because most of these sources would have been detected even without this small amount of boosted radiation from the jet, and their distribution of orientations must then be close to random. In particular, we note that the outer jet in PKS 1510-089 has been found, in the CR model, to be directed at an angle of between 12$\arcdeg-24\arcdeg$ from the line-of-sight \citep{hom02}. This means that its radiation would not be significantly boosted.

Since the sources involved here are radio-loud AGNs found in early surveys made at low frequencies and with large beamwidths, in most cases their detection will have been based on the total flux. Here we find that instead of the boosted radiation representing at least $90\%$ of the total flux, it is the unboosted radiation that is dominant, appearing to represent $\sim90\%$ of the total flux density from many of the radio loud quasars with high $\beta_{app}$ values. This situation will only be worsened if the spectral index of the inner jet is not flat, as assumed here, and actually falls off at the low frequencies of the finding surveys.

It is worth noting that this model, where the jets turn on and off and are closer to the plane of the sky, would then easily explain why \citet{hom02} found no evidence for a counter-jet in PKS 1510-089 by simply interpreting the arcsec jet, lying $\sim180\arcdeg$ from the milliarcsec jet, as the counter-jet. This would require, as is being proposed here for most of these sources, that the jet and counter-jet are both at large viewing angles instead of being viewed end-on as proposed by \citet{hom02}. In this scenario there is also evidence that the jets in PKS 1510-089 switch on and off, as is required to explain intrinsic one-sidedness. Furthermore, as noted above, the polarization detected at the end of the mas jet is also consistent with this scenario. Although in the CR model the superluminal motion of the blobs in the pc-scale jet of 3C279 requires a viewing angle of $i$ = 2$\arcdeg$ to explain \citep{hov09a}, our results indicate that the viewing angle of $30\arcdeg$ to $40\arcdeg$ found for the inner jet by \citet{car93} may actually be the correct one.

\begin{figure}[t]
\hspace{-0.5cm}
\vspace{-1.0cm}
\epsscale{1.0}
\includegraphics[width=8cm]{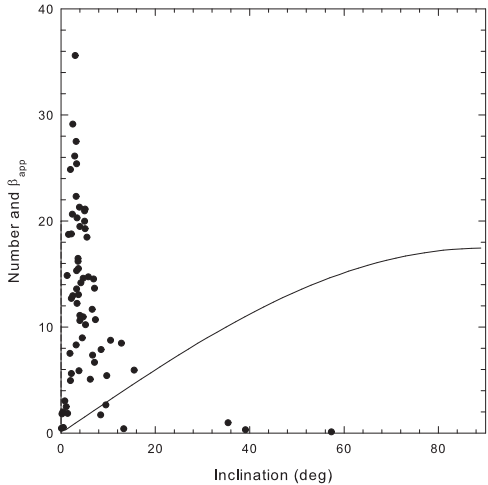} 
\caption{{(solid line) The sin$i$ number vs jet viewing angle distribution expected if no selection effects are active in source finding surveys and 500 sources are found. (points)Actual $\beta_{app}$ vs jet viewing angle distribution from \citet{hov09a}, assuming quasar redshifts are cosmological. \label{fig2}}}
\end{figure}

In summary, when the strengths of the boosted and unboosted radiation are compared, only a very small percentage of the total flux density of these radio loud quasars can be coming from material that is being ejected in a tightly confined beam and at relativistic speeds, and it must be concluded that \em Doppler boosting is unlikely to have played a significant role in the finding surveys in which radio-loud quasars were detected. \em

Astronomers have been aware of this problem ever since the relativistic beaming model was first proposed to explain superluminal motion. At that time there were some concerns that it might be difficult to explain the large number of jets with small viewing angles that seemed to be required, and it was partly this concern that \citet{sch87} was expressing when he stated that \em it is the theoretician's duty to look for ways of escape if the observations should confound the predictions. \em For a review see \em Superluminal Radio Sources, \em ed. Zensus and Pearson, Cambridge University Press; \em Parsec-Scale Radio Jets, \em ed. Zensus and Pearson, Cambridge University Press.

\section{Discussion}

If there are no selection effects operating the distribution of orientations for these sources will be given by the well-known sin$i$ curve in Fig 2 (represented by the solid curve), which is the curve predicted for a random distribution of viewing angles. In Fig 2 the vertical axis represents the number of sources expected in each 2 degree-wide inclination bin, for inclinations between $0\arcdeg$ and $90\arcdeg$, assuming the finding survey found a total of 500 sources. By summing sources at 2, 4, and 6 degrees, it is found that only 6 ($1\%$) of the detected sources would have had inclination angles that are close to the line of sight (below $8\arcdeg$). Also included in Fig 2 (circular points) are $\beta_{app}$ values calculated for sources studied by \citet{kel98,kel04}. These are plotted vs viewing angle on the same scale. In this case the jet viewing angles are those required in the CR model to explain the measured $\beta_{app}$-values, as calculated by \citet{hov09a,hov09b} for 67 jetted sources. In this plot 56 of the 67 sources, or $84\%$, require jet viewing angles $i < 8\arcdeg$. It also shows that if the redshifts are cosmological almost all sources with $\beta_{app} > 5$ are required to have viewing angles $i < 8\arcdeg$, whereas we have just shown that almost none can fall in this category because only a very small percentage of the flux can be boosted resulting in a close-to-random distribution. It is also worth noting that even if the predicted number distribution curve were flat, or even cos$i$, it would be impossible to reconcile it with the observed number distribution obtained using $\beta_{app}$ in Fig 2. 
 
\begin{figure}[t]
\hspace{0.0cm}
\vspace{0.0cm}
\epsscale{1.0}
\includegraphics[width=7cm]{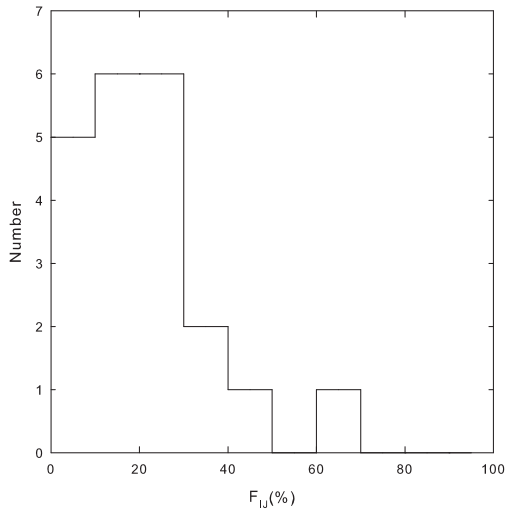}
\caption{{Number of sources from Table 1 plotted versus F$_{IJ}$, the percentage of the flux in the inner jet compared to the total flux in the finding survey. \label{fig3}}}
\end{figure}



Although it has been assumed in Fig 2 that 500 sources above 1 Jy would be found, this number may be too high for radio-loud quasars. The Wall and Peacock sample \citep{wal85} contains 233 bright extragalactic radio sources found in the major centimeter-wavelength surveys at Parkes, Green Bank, and Bonn, and is complete to 2 Jy at 2.7 GHz. The list of bright radio sources at 178 MHz \citep{lai83} contains 181 sources. The K$\ddot{u}$hr sample \citep{kuh81,sti94} contains 518 sources and is complete to 1 Jy at 5 GHz. The 3CR sample has a similar number. However, each of these samples contains many mature radio galaxies that are not part of the high-redshift quasar sample considered here. For example, the K$\ddot{u}$hr sample contains 165 radio galaxies \citep[see Fig 1]{bel07a}. The nature of the redshifts of these mature radio galaxies is not in question and is assumed to be cosmological. The complete radio-loud quasar sample (i.e. quasars and BL Lacs) can therefore be assumed to contain closer to 330 sources, which is considerably less than 500.


The 117 radio-loud quasars and BL Lacs with jets included in the \citet{kel98,kel04} sample thus represent many of those found mainly in the early surveys, which would have been found because of their strong, total flux density, almost all of which is un-boosted. Therefore, for a random distribution, \em less than $\sim3$ of these sources would be expected to have jet viewing angles less than $8\arcdeg$. \em In that study $86\%$ of the sources had $\beta_{app} > 1$, $63\%$ had $\beta_{app} > 3$, $50\%$ had $\beta_{app} > 5$. There were 16 3C-sources in their sample and $50\%$ of these had $\beta_{app} > 3$. There were 21 4C-sources and, of these, $80\%$ had $\beta_{app} > 3$ and $65\%$ had $\beta_{app} > 5$. Overall, there were 34 sources, or $26\%$, with $\beta_{app} > 10$ ($i < 8\arcdeg$). Of these, 25 are PKS or 4C sources, or both. Almost all of these (23) are high redshift sources, with redshifts greater than $z$ = 0.6.

In the K$\ddot{u}$hr sample, approximately $75\%$ of the 269 quasars with measured redshifts ($\sim200$ sources) and measured spectral index, have reasonably flat spectra. If $26\%$ of these have $\beta_{app} > 10$, like those in the \citet{kel04} sample, then $\sim50$ of these would have to have viewing angles less than $8\arcdeg$, where less than $\sim3$ are expected for a random distribution. The Kellermann sample was drawn from the list of radio-loud sources found in the original surveys and if these lists contain only a very few sources with small viewing angles, no matter how the sources are later chosen \em it cannot change the total number with small viewing angles that are available to be chosen. \em Since the evidence then indicates that almost none of the sources with high-$\beta_{app}$ values can have been preferentially selected because of Doppler boosting, almost all must have viewing angles $i > 8\arcdeg$.

The results found here also mean that the term $blazar$ needs to be more clearly defined. This term has come to represent a quasar or BL Lac object whose variability results mainly from the fact that the jet is pointed in our direction \citep{kha10}. It now can imply only that the flux density fluctuations seen in AGNs are due simply to the fact that the central engine is currently swallowing, and spitting out, new in-falling material, without any implication that the jet is pointed in our direction. This explanation also fits the observations better since the fluctuations in 3C279 \citep{bel10,cha08} and other radio variable sources are observed to be associated mostly with the growth and decline of the $number$ of blobs moving away from the core at any given time, and not with simultaneous fluctuations in all blobs. The latter might be expected if, as has been previously suggested, the fluctuations were due to small changes in the viewing angle of jets closely aligned with the line-of-sight. The fact that this is not seen also agrees with our finding that the jet viewing angles are large in almost every case.

\subsection{How Complete is the Radio-Loud AGN Sample?}

If, for example, we assume that 95 percent of the radio radiation from radio-loud quasars is coming from material that is moving out in a jet at relativistic speeds, then because of the relativistic beaming effect where the radiation is enhanced in the direction of motion, those sources with their jets pointed in our direction would be significantly stronger than those whose jets have large viewing angles. In detection-limited finding surveys many of the sources whose jets have large viewing angles would then fall below the detection limit while those pointed in our direction would be detected. In this scenario the high percentage of radio-loud quasars requiring small viewing angles could be explained as representing that few percent of sources in a random sample that have small viewing angles, while the remaining $\sim95\%$ of the sample lies below the detection limit. However, if, on the other hand, only a small percentage of the source radiation comes from material that is moving out at relativistic speeds, the flux from those sources with large jet viewing angles would not differ significantly from those with small viewing angles. In this scenario almost all of the radio-loud sources would be detected and the sample would be essentially complete. We have shown above that it is this latter situation that is most likely to be the correct one.

In Fig 3, the number of sources from Table 1 is plotted versus F$_{IJ}$, the relative percentage of boosted to un-boosted flux. The number peaks near F$_{IJ}$ = 20$\%$. If the boosted inner jet is only 20$\%$ of the total flux, the strengths of radio-loud AGNs with larger viewing angles would not be expected to be significantly fainter than those with small viewing angles even if the entire inner jet disappeared. The radio-loud sample would then be expected to be reasonably complete above 1 Jy for all viewing angles. It is concluded that the 330 radio-loud quasars in the K$\ddot{u}$hr sample make up close to a complete radio-loud sample. This makes the high percentage of sources observed with large apparent superluminal motions very difficult to explain statistically. 
 
In summary, when the Doppler boosted component is small compared to the total source flux it can be concluded that the sample of radio-loud sources detected in a finding survey will represent almost all of the radio-loud sources and only a few percent of them can have small viewing angles. In this situation some explanation other than relativistic beaming must be found to explain the high percentage of sources exhibiting large apparent superluminal motions. 
 
\subsection{Other Radio Selection Effects}

There may still be some radio selection effects present that are related to viewing angle but unrelated to Doppler boosting. For example if there is a torus surrounding the central object it can block some of the radiation coming from the central compact object when inclinations are large (near edge-on). Recently \citet{lov10,lov11} have measured the opening angles and inclinations for 55 radio quiet quasars. They found opening angles near 78$\arcdeg$ in these objects and the distribution of inclinations has been plotted in Fig 4, where it can be seen that for small inclination angles the number distribution follows the sin$i$ curve closely. The vertical dashed line indicates the inclination angle (78$\arcdeg$/2) above which the torus prevents the central compact object from being viewed. In fact, when viewed in this manner, the results in Fig 4 suggest that the opening angle may be slightly smaller (60$\arcdeg$ - 65$\arcdeg$) than the 78$\arcdeg$ reported by \citet{lov10}. For larger inclination angles the torus has clearly affected the detection of these radio-quiet objects. However, it is obvious from Fig 4 that the radio-quiet distribution does follow a sin$i$ curve for viewing angles that are unaffected by the opening angle cut-off. 

\begin{figure}[t]
\hspace{-1.0cm}
\vspace{-1.0cm}
\epsscale{1.0}
\includegraphics[width=9cm]{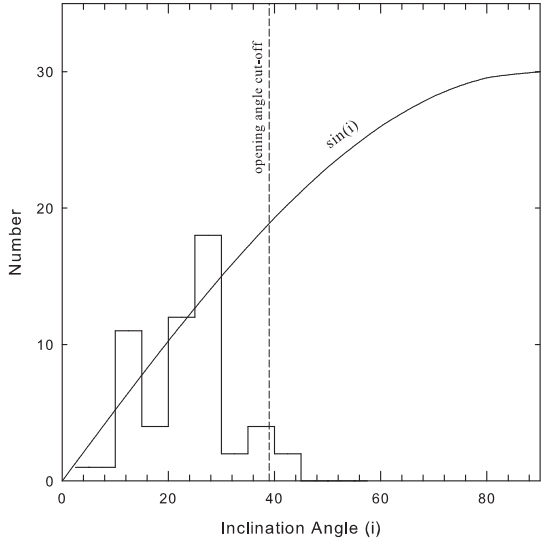}
\caption{{Viewing angle distribution for 55 radio-quiet quasars from \citet{lov10}. The vertical dashed line indicates the viewing angle above which the opening angle of 78$\arcdeg$ found by these authors would be expected to affect the number distribution. \label{fig4}}}
\end{figure}

\subsection{Radio Quiet Objects as the Parent Population of Radio Loud Objects}

There is clear confirmation from Fig 4 that without a dominant Doppler boosted component present few sources with viewing angles less than 10º will be detected. If the radio-quiet quasars really represent the parent sample from which the radio-loud quasars are drawn, when the radio-loud quasars contain only a small boosted component, as found here, the radio-loud sources must then be radio loud because they are closer, but they are still expected to have the same sini distribution of orientations as given by the solid curve in Fig 4. The radio-loud distribution given by the filled circles in Fig 2, and determined assuming that the distance to these objects is reliably given by their redshifts, is clearly incompatible with the sini distribution found for the radio-quiet sources in Fig 4. This is strong confirmation that if the Doppler boosted component is small, as found here, the redshifts of the objects in Fig 2 cannot be an accurate indication of distance.



\begin{deluxetable}{cccccccccc}
\tabletypesize{\scriptsize}
\tablecaption{Percent of Flux Likely to be Boosted for High-$\beta_{app}$ Sources. \label{tbl-1}}

\tablewidth{0pt}
\tablehead{

\colhead{Source} & \colhead{Alt. name} & \colhead{$\beta_{app}$} & \colhead{z} & \colhead{S$_{178}$(Jy)} & \colhead{S$_{400}$(Jy)} & \colhead{S$_{in}$(Jy)\tablenotemark{a}} & \colhead{S$_{ext}$(Jy)\tablenotemark{b}} & \colhead{F$_{IJ}(\%)$\tablenotemark{c}} & \colhead{F$_{EJ}(\%) $\tablenotemark{d}}
}

\startdata

0106+013 & PKS & 23 & 2.1   & --- & 3.5  & 0.33 & 0.531 & 9.4 & 24.6  \\
0149+218 & PKS & 18 & 1.32  & --- & 1.9  & 0.16 & 0.025 & 8.4 & 9.7 \\
0234+285 & 4C+28.07 & 13 & 1.213 & 2.1 & ---  & 0.49 & 0.1  & 23 & 28.1 \\
0333+321 & 4C+32.14 & 24 & 1.263 & 2.2 & --- & 0.4  & 0.072 & 18.2 & 21.6 \\
0420-014 & PKS & 14 & 0.92  & 1.2 & 1.5  & 0.64 & 0.070 & 42 & 47 \\
0850+581 & 4C+58.17 & 13 & 1.32  & 2.9 & ---  & 0.24 & ---  & 8.3 & 8.3 \\
0945+408 & 4C+40.24 & 22 & 1.252 & 2.5 & ---  & 0.56 & 0.095 & 22 & 26 \\
1156+295 & 4C+29.45 & 8.9 & 0.729 & 2.8 & --- & 0.34 & 0.196 & 12 & 19.1 \\
1226+023 & 3C 273 & 14 & 0.158 & 75 & --- & 16.5 & 17.6 & 22 & 45 \\ 
1508-055 & PKS & 31 & 1.18  & --- & 8.9  & 0.19 & ---  & 2.1 & 2.1 \\
1510-089 & PKS & 19 & 0.36  & --- & 3.0  & 0.46 & 0.18 & 15 & 21.3 \\
1606+106 & 4C+10.45 & 30 & 1.226 & 2.7 & 4.4  & 0.33 & 0.26 & 12 & 22 \\
1633+382 & 4C+38.41 & 11.5 & 1.807 & 2.2 & --- & 0.67 & 0.032 & 30 & 32 \\
1641+399 & 3C345 & 17 & 0.594 & 10  & ---  & 3.95 & 1.48 & 39 & 54 \\
1642+690 & 4C+69.21 & 16 & 0.751 & 2.5 & ---  & 0.27 & 0.33 & 10 & 24 \\
1730-130 & PKS & 23 & 0.90  & --- & 6.3  & 1.09 & 0.517 & 17 & 25.5 \\
1823+568 & 4C+56.27 & 3.4 & 0.663 & 2.4 & --- & 0.26 & 0.137 & 10.8 & 16 \\
1828+487 & 3C380 & 15 & 0.692 & 57 & ---  & 1.0  & 5.43 & 1.7 & 11 \\
2201+315 & 4C+31.63 & 6.3 & 0.298 & 3.5 & --- & 0.82 & 0.378 & 23 & 34 \\ 
2223-052 & 3C446 & 32 & 1.404 & 17.3 & --- & 0.64 & 0.92 & 4 & 9 \\

\enddata 

\tablenotetext{a}{Flux in the Inner Jet at 2 cm where S$_{in}$ = S$_{total}$ - S$_{peak}$ from \citet{kel98}}.
\tablenotetext{b}{Flux in external (kpc) jet from \citep{kha10,mur93}}.
\tablenotetext{c}{F$_{IJ}$ = Percentage of Inner Jet flux (S$_{in}$)compared to total flux in finding survey. Assumes a flat jet spectral index.}
\tablenotetext{d}{F$_{EJ}$ = Percentage of Entire (inner and outer) jet flux compared to total flux in finding survey. Assumes a flat spectral index.}

\end{deluxetable}

\section{Conclusions}  

It is concluded here that Doppler boosting could not have played a significant role in finding radio-loud, high redshift quasars because the component of their radiation that comes from material being ejected outwards at relativistic speeds, and in a tightly confined jet, is insignificant compared to the total flux obtained in the low-frequency finding surveys. This is true even if the radiation from the kpc-scale jet is from material that is moving relativistically, and is especially true for sources with large radio lobes. Without a highly directed, relativistically beamed component that dominates the source flux density, sources cannot be preferentially selected with small jet viewing angles and the resulting distribution of jet viewing angles will then be close to that of a random one (sin$i$). In this case almost all will have viewing angles much greater than $8\arcdeg$ and even the flux from the inner jet will be un-boosted. This means that relativistic ejections with small jet viewing angles cannot be used to explain the observed superluminal motions seen in high-redshift quasars. Although this problem can easily be resolved by bringing these sources closer and accepting intrinsic redshift components in high redshift quasars, this solution has so far been found unacceptable by most astronomers. At the very least, a new explanation for superluminal motion that does not involve relativistic beaming will need to be found if the redshifts of high redshift quasars are to remain cosmological.

\acknowledgements

I thank S. Comeau and D. McDiarmid for helpful comments when this manuscript was being prepared.


\clearpage

\begin{thebibliography}{}


\bibitem[Alexander and Leahy(1987)]{ale87} Alexander, P. and Leahy, J.P. 1987, \mnras, 225, 1
\bibitem[Bell(2007a)]{bel07a} Bell, M.B. 2007a, \apjl, 667, L129 (astro-ph/0704.1631)
\bibitem[Bell(2007b)]{bel07b} Bell, M.B. 2007b, astro-ph/0711.4531

\bibitem[Bell and Comeau (2010)]{bel10} Bell, M.B., and Comeau, S.P. 2010, \apss, 325, 31 (DOI: 10.1007/s10509-009-0162-z)

\bibitem[Biretta et al.(1999)]{bir99} Biretta, J.A., Sparks, W.B., and Macchetto, F. 1999, \apj, 520, 621
\bibitem[Bolton et al.(1964)]{bol64} Bolton, J.G., Gardner, F.F. and Mackey, M.B. 1964, AJP, 17, 340
\bibitem[Carrara et al.(1993)]{car93} Carrara, E.A., Abraham, Z., Unwin, S.C., and Zensus, J.A. 1993, \aap, 279, 83  
\bibitem[Chatterjee et al.(2008)]{cha08} Chatterjee, R. et al. 2008, \apj, 689, 79
\bibitem[Cleary et al.(2006)]{cle06} Cleary, K., Lawrence, C.R., Marshall, J.A., Hao, L., and Meier, D. 2007, \apj, 660, 117
\bibitem[Dixon(1970)]{dix70} Dixon, R.S., 1970, \apjs, 20, 1
\bibitem[Ekers(1969)]{eke69} Ekers, J. 1969, Aust. J. Phys. Suppl., 7, 3
\bibitem[Gower et al.(1967)]{gow67} Gower, J., Scott, P. and Wills, D. 1967, \memras, 71, 49
\bibitem[Homan et al.(2002)]{hom02} Homan, D.C., Wardle, J.F.C., Cheung, C.C., Roberts,, D.H., and Attridge, J.M. 2002, \apj, 580, 742
\bibitem[Hovatta et al.(2009a)]{hov09a} Hovatta, T., Valtaoja, E., Tornikoski, M., and Lahteenmaki, A. 2009a, \aap, 494, 527
\bibitem[Hovatta et al.(2009b)]{hov09b} Hovatta, T., Valtaoja, E., Tornikoski, M., and Lahteenmaki, A. 2009b, \aap, 498, 723
\bibitem[Jorstad et al.(2010)]{jor10} Jorstad, S.G. et al. 2010, preprint in arXiv:1003.4293 [astro-ph.CO]

\bibitem[Kellermann et al.(1998)]{kel98} Kellermann, K.I., Vermeulen, R.C., Zenus, J.A., and Cohen, M.H. 1998, \aj, 115, 1295
\bibitem[Kellermann et al.(2004)]{kel04} Kellermann, K.I. et al. 2004, \apj, 609, 539
\bibitem[Kharb et al.(2010)]{kha10} Kharb, P., Lister, M.L., and Cooper, N.J. 2010, preprint in astro-ph/1001.0731
\bibitem[Kovalev et al.(2007)]{kov07} Kovalev, Y.Y., Lister, M.L., Homan, D.C. and Kellermann, K.I. 2007, \apjl, 668, L27
\bibitem[Kuhr et al.(1981)]{kuh81} K$\ddot{u}$hr, H., Witzel, A., Pauliny-Toth, I.I.K., and Nauber, U. 1981, \aaps, 45, 367
\bibitem[Laing and Bridle(2008)]{lai08} Laing, R.A., and Bridle, A.H. 2008, ASP Conference Series 386 (Extragalactic Jets: Theory and Observation from Radio to Gamma Ray), T.A. Rector and D.S. Yong eds. (arXiv/0801.0147) 
\bibitem[Laing et al.(1999)]{lai99} Laing, R.A., Parma, P., De Ruiter, H.R., and Fanti, R. 1999, \mnras, 306, 513
\bibitem[Laing and Bridle(2002a)]{lai02a} Laing, R.A. and Bridle, A.H. 2002, \mnras, 336, 328
\bibitem[Laing and Bridle(2002b)]{lai02b} Laing, R.A. and Bridle, A.H. 2002, \mnras, 336, 1161
\bibitem[Laing et al.(2006)]{lai06} Laing, R.A., Calvin, J.R., Bridle, A.H., and Hardcastle, M.J. 2006, \mnras, 372, 510
\bibitem[Laing et al.(1983)]{lai83} Laing, R.A., Riley, J.M., and Longair, M.S. 1983, MNRAS, 204, 151 
\bibitem[Lind and Blandford(1985)]{lin85} Lind, K.R. and Blandford, R.D. 1985, \apj, 295, 358
\bibitem[L$\acute{o}$pez-Corredoira(2011)]{lop11} L$\acute{o}$pez-Corredoira, M. 2011, International J. of Astron. and Astrophys., 1, 73
\bibitem[Lovegrove et al.(2010)]{lov10} Lovegrove, J., Schild, R.E., and Leiter, D. 2010, (preprint in arxiv:1003.5497)
\bibitem[Lovegrove et al.(2011)]{lov11} Lovegrove, J., Schild, R.E., and Leiter, D. 2011, \mnras, 412, 2631
\bibitem[Marscher(1977)]{mar77} Marscher, A.P. 1977, \apj, 216, 244
\bibitem[Marscher(2008)]{mar08} Marscher, A.P. 2008, in \em Extragalactic Jets: Theory and Observations from Radio to Gamma Rays, \em ASP Conference Series, Vol 386, eds. T.A. Rector and and D.S. De Young
\bibitem[Marscher(2009)]{mar09} Marscher, A.P. 2009, arXiv:0909.2576v1
\bibitem[Miley(1971)]{mil71} Miley, G.K. 1971, MNRAS, 152, 477
 
\bibitem[Murphy et al.(1993)]{mur93} Murphy, D., Browne, I., and Perley, R. 1993, MNRAS, 264, 298

\bibitem[Narlikar(1989)]{nar89} Narlikar, J.V. 1989, \ssr, 50, 523
\bibitem[O$\arcmin$Dea et al.(2009)]{ode09} O$\arcmin$Dea  et al. 2009, \aap, 494, 471
\bibitem[Pilkington and Scott(1964)]{pil64} Pilkington, J. and Scott, P. 1964, \memras, 69, 183
\bibitem[Rees(1966)]{ree66} Rees, M.J. 1966, Nature, 211, 468
\bibitem[Scheuer(1987)]{sch87} Scheuer, P. in \em Superluminal Radio Sources \em eds Zensus, J.A. and Pearson, T.J. 1987, p104, (Cambridge:Cambridge University Press)
\bibitem[Sikora et al.(2008)]{sik08} Sikora, M, Moderski, R., and Madejski, G. M. 2008, \apj, 675, 71
\bibitem[Stickel et al.(1994)]{sti94} Stickel, M., Meisenheimer, K., and Kuhr, H. 1994, \aaps, 105, 211
\bibitem[Urry and Padovani(1995)]{urr95} Urry, C.M., and Padovani, P. 1995, \pasp, 107, 803
\bibitem[Wall and Peacock(1985)]{wal85} Wall, J.V. and Peacock, J.A. 1985, MNRAS, 216,173
\bibitem[Zensus and Pearson(1987)]{zen87} Zensus, J.A. and Pearson, T.J. 1987, in \em Superluminal Radio Sources, \em  eds. Zensus and Pearson, 1987, Cambridge University Press

\end{thebibliography}
\end{document}